\documentclass{aa}  

\bibpunct{(}{)}{;}{a}{}{,} 
\usepackage{natbib}
\usepackage{graphicx}
\usepackage{txfonts}
\usepackage{lscape}
\usepackage{color}
\begin{document} 

\title{Dissecting the AGB star L$_{\rm 2}$ Puppis: a torus in the making \thanks{Based on observations from ESO programs 090.D-0541 \& 090.D-0677}}

   \author{
     F. Lykou\inst{1}
          \and D. Klotz\inst{1}
          \and C. Paladini\inst{1,2}
          \and J. Hron\inst{1}
          \and A. A. Zijlstra\inst{3}
          \and J. Kluska\inst{4}
          \and B. R. M. Norris\inst{5}
          \and P. G. Tuthill\inst{5}
          \and S. Ramstedt\inst{6}
          \and E. Lagadec\inst{7,8}  
          \and M. Wittkowski\inst{9}
	\and M. Maercker\inst{10}
          \and A. Mayer\inst{1}        
}

   \institute{Institut f\"ur Astrophysik, Universit\"at Wien, T\"urkenschanzstrasse 17, 1180, Wien, Austria
         \and Institut d'Astronomie et d'Astrophysique, Universit\'e Libre de Bruxelles, CP 226, Boulevard du Triomphe, 1050, Bruxelles, Belgium
         \and Jodrell Bank Centre of Astrophysics, Alan Turing Building, University of Manchester, Oxford Road, M13 9PL, Manchester, UK
         \and UJF-Grenoble 1/CNRS-INSU, Institut de Plan\'etologie et d'Astrophysique de Grenoble (IPAG), UMR, 5274, France 
          \and Sydney Institute for Astronomy, University of Sydney, 44 Rosehill St, Redfern, NSW 2016, Sydney, Australia
         \and Department of Astronomy, Uppsala Universitet, Box 516, 75120 Uppsala, Sweden
         \and Department of Astronomy, Cornell University, 222 Space Sciences Building, Ithaca, NY 14853-6801, USA
         \and Laboratoire Lagrange, UMR 7293, Universit\'e de Nice-Sophia Antipolis, CNRS, Observatoire de la C\^ote d'Azur, Bd de l'Observatoire, B.P. 4229, F-06304 Nice cedex 4, France
         \and European Southern Observatory, Karl-Schwarzschild-Strasse 2, 85748 Garching, Germany
          \and Onsala Space Observatory, Department of Earth and Space Sciences, Chalmers University of Technology, SE439-92 Onsala, Sweden
                             }

   \date{Received October 10, 2013; accepted Month 00, 0000}

 
  \abstract
   {}
   {The circumstellar environment of \object{L$_{\rm 2}$ Pup}, an oxygen-rich semiregular variable, was observed to understand the evolution of mass loss and the shaping of ejecta in the late stages of stellar evolution.}
   {High-angular resolution observations from a single 8 m telescope were obtained using aperture masking in the near-infrared (1.64, 2.30 and 3.74~$\rm\mu m$) on the NACO/VLT, both in imaging and polarimetric modes.}
   { The aperture-masking images of L$_{\rm 2}$ Pup at 2.30~$\rm\mu m$ show a resolved structure that resembles a toroidal structure with a major axis of $\sim$140 milliarcseconds (mas) and an east--west orientation. Two clumps can be seen on either side of the star, $\sim$65  mas from the star, beyond the edge of the circumstellar envelope (estimated diameter is $\sim$27~mas), while a faint, hook-like structure appear toward the northeast. The patterns are visible both in the imaging and polarimetric mode, although the latter was only used to measure the total intensity (Stokes $I$). The overall shape of the structure is similar at the 3.74~$\rm\mu m$ pseudo-continuum (dust emission), where the clumps appear to be embedded within a dark, dusty lane. The faint, hook-like patterns are also seen at this wavelength, extending northeast and southwest with the central, dark lane being an apparent axis of symmetry.  We interpret the structure as a circumstellar torus with inner radius of 4.2\,au. With a rotation velocity of 10 km s$^{-1}$ as suggested by the SiO maser profile, we estimate a stellar mass of 0.7\,M$_\odot$.}
   {}


   \maketitle
%

\section{Introduction}

During the late stages of evolution of intermediate-mass stars (1--8~M$_{\sun}$) -- that is, at the transition from the asymptotic giant branch (AGB) to the post-AGB phase -- the ejected material is shaped into intriguing forms (e.g., bipolar nebulae, torii, disks and/or spirals) that depart from the initial spherical symmetry~\citep[][ and references therein]{vanwinckel2003}. Although there are many studies in the literature of asymmetries found in post-AGB stars, detections of such asymmetries in AGB stars are rare. It is therefore uncertain at which point the asymmetries begin to develop.

One of the brightest AGB stars in the near-infrared ($m_{\rm K}\leq-2.1$~mag) near the solar neighborhood \citep[$64\pm4$~pc; ][]{vanleeuwen2007} is \object{L$_{\rm 2}$ Pup} (HD56096, HIP34922, IRAS 07120-4433). It is a semiregular variable (M5III; $P=140\pm1$ days) that experienced a dimming event starting around 1995 ($\rm\Delta V\sim$3 mag; Fig.~\ref{fig:lightcurve}), which was attributed to circumstellar dust~\citep{bedding2002}. The decline lasted for six years, and as of June 2014, the star has not yet begun to recover its former brightness. Dimming events are present in a few other evolved stars~\citep[e.g., V Hya; ][]{knapp1999}, which may repeat over the course of many years.  L$_{\rm 2}$ Pup had a less severe dimming event around 1960: recovery took until 1980. Unlike V~Hya, there is no clear periodicity to the dimming in the data, which go back to 1930.

Only a few images of L$_{\rm 2}$ Pup are available in the literature.~\citet{bedding2002} stated that the IRAS images are non-descriptive. The WISE satellite observed L$_{\rm 2}$ Pup post-dimming in the near- and mid-infrared at low angular resolutions, although all images were saturated by the stellar emission.~\citet{jura2002} studied L$_{\rm 2}$ Pup with the Keck 10 m telescope in the mid-infrared and inferred that two asymmetries are present at position angles 135\degr\ and 225\degr\ east-of-north. Past studies on the variation of polarization of L$_{\rm 2}$ Pup are indicative of an asymmetric\footnote{In this paper, the term `asymmetric' will describe any aspherical structure.} dust distribution~\citep{magalhaes1986}. In addition to the stellar component, \citet{ireland2004} inferred from the MAPPIT/AAT (3.9 m) observations in the optical that an additional emitting component is required. They probed different spatial scales above the photosphere and found a size variation with respect to wavelength in the titanium oxide band (0.68 -- 0.82~$\rm\mu$m) as well as an asymmetric brightness profile, which they attributed to scattering by dust. The indications of asymmetry in the circumstellar environment of L$_{\rm 2}$ Pup, suggested by polarimetry~\citep{magalhaes1986}, as well as the indications of dust scattering by interferometric observations (\citet{ireland2004}; see also Section~\ref{sec:past}), might indicate a disk or a companion.


\begin{figure*}[!htb]
\begin{minipage}[b]{0.45\linewidth}
	\centering
		\includegraphics[width=0.9\textwidth, bb=49 144 532 640]{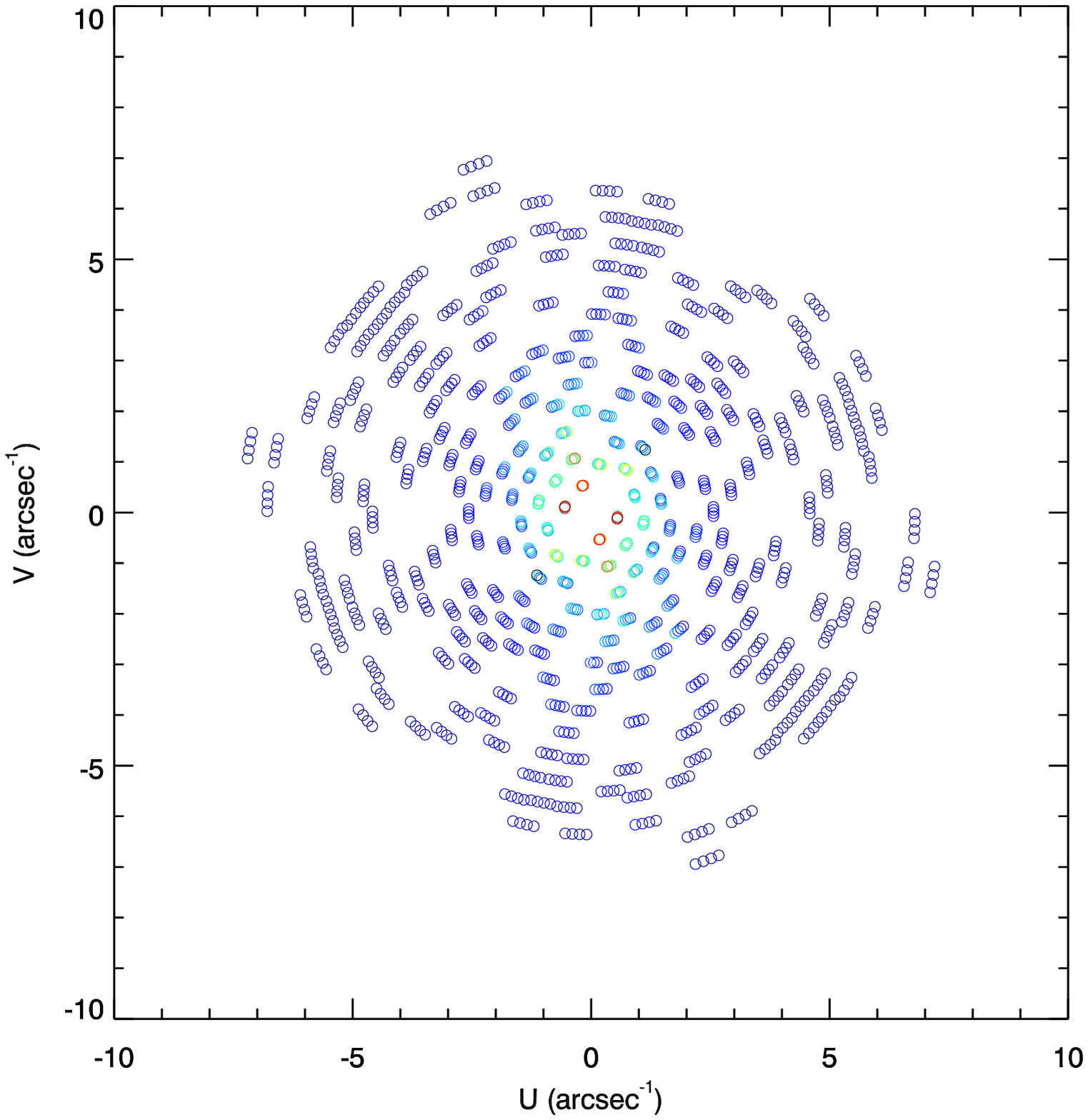}
	\end{minipage}
	\hspace{0.5cm}
	\begin{minipage}[b]{0.45\linewidth}
	\centering
		\includegraphics[width=0.9\textwidth, bb=41 164 524 660]{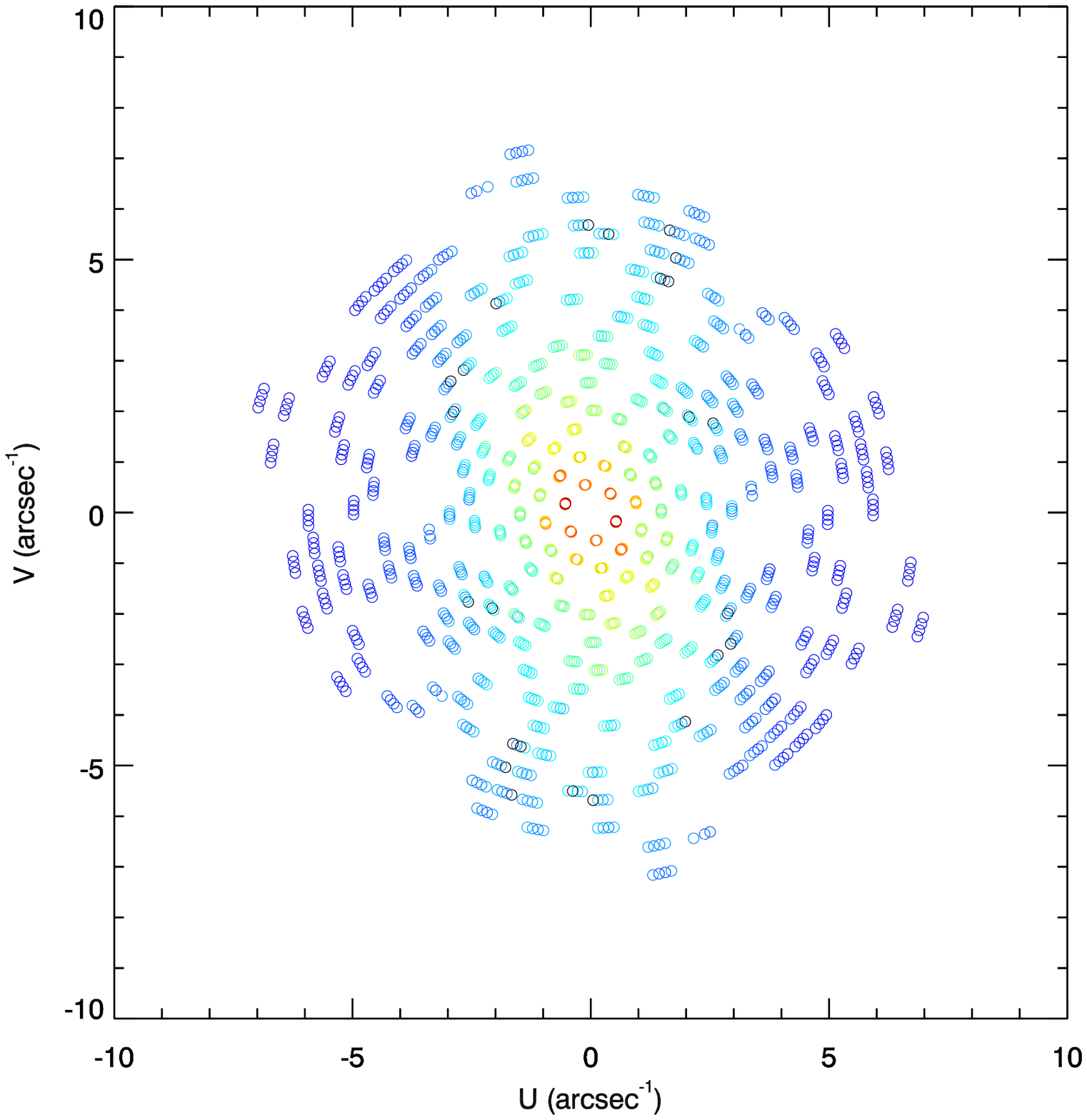}
	\end{minipage}
	\caption{SAM/NACO \textit{uv}-plane coverage with the 18-hole mask for 1 December 2012 at 2.30~$\rm\mu$m (left) and 3.74~$\rm\mu$m (right). Each \textit{uv} point is color-coded to the respective squared visibility value descending gradually from unity (red) to zero (dark blue). }
	\label{fig:uvplane}
\end{figure*}

Very recently, in an effort parallel to our own, \citet{kervella2014} have presented infrared interferometric data and deconvolved adaptive-optics images, which show a disk-like structure, approximately 700 mas across, transecting the star, opaque at $J$, translucent at $H$ and $K$, and a thermal dust emitter at $L$. A faint loop extends to the northeast. Kervella et al. attributed the dimming to this disk. L$_{\rm 2}$ Pup is therefore one of the few AGB stars with evidence for asphericities in their circumstellar material, and as the closest mass-losing AGB star, is ideally placed for resolving the details of the envelope structure. We note that the technique of aperture masking used here is different and complementary to the technique of direct adaptive optics imaging used by \citet{kervella2014}, as discussed in detail in Sect.~\ref{ker}.

The circumstellar environment of L$_{\rm 2}$~Pup was observed in the near-infrared, using aperture masking techniques on an 8.2 m telescope (NACO/VLT), to detect any asymmetries within a 0\farcs5 radius. The observations are presented in the next section, while results are discussed in Sects.~\ref{sec:asymmetry} and~\ref{sec:imaging}. Conclusions are discussed in Sects. \ref{sec:discussion} and \ref{sec:results}. 

\begin{table*}[!ht]
	\centering
	\caption{Observation log.}
\begin{tabular}{cccccccccc}\hline\hline
Mode & Date & Phase & Filter & $\rm\lambda_c$ ($\rm\Delta\lambda$) [$\rm\mu$m] & \# frames & Camera\tablefootmark{a} & $t_{\rm int}$ [sec] & {Seeing\tablefootmark{b}} & Airmass \\ \hline
SAM & 01/12/2012 & 0.358 &NB\_1.64 & 1.64 (0.04) & 1200 & S13  & 0.5 & 1.00 & 1.10 \\
SAM & 01/12/2012 & 0.358 &IB\_2.30 & 2.30 (0.06) & 600 & S13  & 1.0 & 0.72 & 1.09 \\
SAM & 01/12/2012 & 0.358 &NB\_3.74 & 3.74 (0.02) & 800 & L27  & 0.5 & 1.18 & 1.08 \\
SAMPol &08/02/2013 & 0.855 & IB\_2.30 & 2.30 (0.06) & 2000 & S13 &  0.5 & 1.84 & 1.07 \\
\hline
\end{tabular}
\tablefoot{
\tablefoottext{a}{Camera pixel scales are 13.22 and 27.19 mas/pixel for the S13 and L27 cameras, respectively.}
\tablefoottext{b}{In units of arcseconds.}
}
	\label{tab:log}
\end{table*}

\section{Observations and data reduction}\label{sec:observations} 
L$_{\rm 2}$ Pup was observed on 1 December 2012 (Prog.ID: 090.D-0541, PI: D. Klotz) at NACO/VLT \citep{lenzen2003, rousset2003} in sparse aperture masking mode~\citep[SAM;][]{tuthill2010,lacour2011} using the 18--hole\footnote{Although the mask has 18 holes, only 17 holes are used effectively, because one is located at the edge of the pupil and is thus omitted in the data reduction.} mask at 1.64, 2.30 and 3.74~$\rm\mu$m. 

Additional observations were obtained on 8 February 2013 (Prog. ID: 090.D-0677, PI: A.A. Zijlstra) at 2.30~$\rm\mu$m using the SAM polarimetric mode (SAMPol). In this mode, a Wollaston prism positioned beyond the aperture mask splits the incoming beam into two polarizations \citep[for a detailed description, see][]{norris2012}. With the use of a half-wave plate, the plane of polarization of the beam is rotated at 0\degr, 22.5\degr, 45\degr\ , and 67.5\degr\ to obtain the polarized intensities at 0\degr, 45\degr, 90\degr\ , and 135\degr\ , that is $I_{\rm 0},\,I_{\rm 45},\,I_{\rm 90}$ and $I_{\rm 135}$. The Stokes vector component, $I$, of the polarized beam can be recovered from these intensities through the following relation of \citet{murakawa2005}:
\begin{equation}
I\,=\,\frac{1}{2T}\left( I_{\rm 0} + I_{\rm 45} + I_{\rm 90} + I_{\rm 135} \right),
\end{equation}
\noindent where $T$ is the transmission of the Wollaston prism ($\sim$93\% at 2.30~$\rm\mu$m). 

Table~\ref{tab:log} contains details of each observing run, such as stellar phases, integration times, number of acquisition frames, and seeing conditions. The projected 136 baseline lengths were in the range of 0.5 and 7.8 meters, which within only 10 minutes of observing time provides an excellent \textit{uv}-plane coverage (Fig.~\ref{fig:uvplane}). SAM non-redundant masks allow recovering diffraction-limited images with a relative resolution of 20, 30, and 50~mas at 1.64, 2.30, and 3.74~$\rm\mu m$, respectively, with minimal disturbance by atmospheric seeing.

Pupil-tracking mode was enabled to prevent the apertures from being covered by the telescope's spiders. This leads to field rotation on the detector, indicated by the rotation of the $uv$-plane coverage as seen in Fig.~\ref{fig:uvplane}. All speckle frames were recorded in cube mode. To decrease readout time and the volume of recorded data, two subframes were chosen on the detector: $1024\times1028$ pixels for the S13 camera and $512\times514$ pixels for the L27 camera. Different dither positions in all four quadrants of those subframes were used in a clockwise pattern to minimize readout time and detector noise (due to the variation of photon and thermal noise in each quadrant during data acquisition), as well as simultaneously recording sky background at an opposing dither position. To allow for determining the transfer function and calibration of the observables, $\pi$~Pup ({\tt K3Ib}) was observed immediately after the science target.

The data sets were processed with a pipeline developed at the University of Sydney \citep{tuthill2000,lacour2011}. The interferograms in each data cube were flat-fielded, dark, bias, and sky subtracted. Each speckle frame was Fourier transformed, and the extracted Fourier components (power spectra, bispectra) were calibrated with respect to the unresolved target. Because of calibration problems, the results from the 1.64~$\rm\mu m$ data are inconclusive.

\begin{figure}[!ht]
\centering
       \includegraphics[width=0.495\textwidth, bb=48 41 527 687]{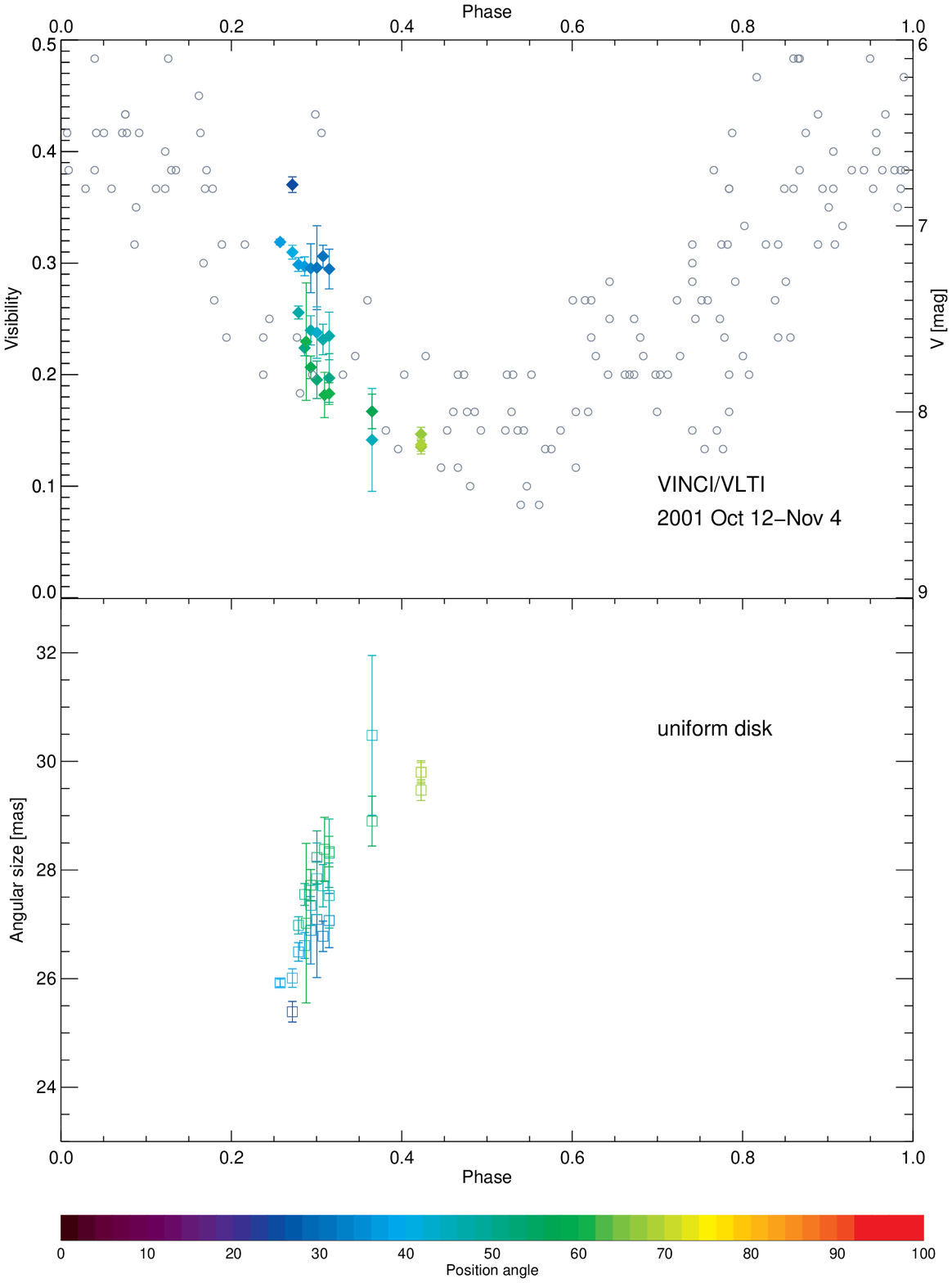}
\caption{VINCI/VLTI observations of L$_{\rm 2}$ Pup from 12 October 2001 to 4 November 2001 with respect to the stellar pulsation phase. The variability of the visibility is shown in the \textit{top panel}, color-coded to indicate its dependency on the baseline position angle (25\degr --70\degr), while the respective uniform disk diameters are shown in the \textit{bottom panel}. A folded visual light curve (open circles) is also shown for comparison in the {\it top panel}. The projected baseline length varies also by less than 1.5 meters (14.66--15.98 m) with respect to the physical baseline length (16 m).}%
\label{fig:vinciplot}
\end{figure}

\section{Fitting the visibilities: indications of an extended component}\label{sec:asymmetry}

\subsection{Past interferometric observations}\label{sec:past}

\citet{ochsenbein1982} predicted a diameter for L$_{\rm 2}$ Pup in the visual of 13.18~milliarcseconds (mas) based on calculations from the intrinsic color and brightness. \citet{ireland2004} previously studied L$_{\rm 2}$ Pup with optical interferometric techniques using aperture masking, although their projected baselines were oriented at 180$^o$, that is north--south. They measured an extended component whose size varies with wavelength (Gaussian full-width-half-maximum $76\pm6$~mas), and the diameter of a compact source (uniform disk, UD) at $25\pm2$~mas ($\sim$50\% of the total flux), which is almost twice the size of the theoretical prediction of \citet{ochsenbein1982}.

VINCI/VLTI data \citep{meisner2008} show a similar size for the \textit{K}-band emission ($2.2\pm0.2\,\rm\mu m$; $\bar{\theta}_{\rm UD}=28^{+4}_{-3}$~mas). This VINCI diameter varies over less than a month of observations, but the visibility is dependent on the different projected baselines and on the stellar pulsation phase, that is the duration of the VINCI observations spanned over one fifth of the stellar pulsation period, as shown in Fig.~\ref{fig:vinciplot}.  \citet{kervella2014} also found evidence for an unresolved component with an upper limit of $\theta_{\rm UD}=17.5\pm1.6$~mas, which they interpreted as the central star.  The resolved components (76 and 28 mas; MAPPIT) would be affected by circumstellar emission.

\subsection{This work}

\begin{figure}[!hbtp]
  \centering
  \includegraphics[width=0.49\textwidth, bb=36 107 546 757]{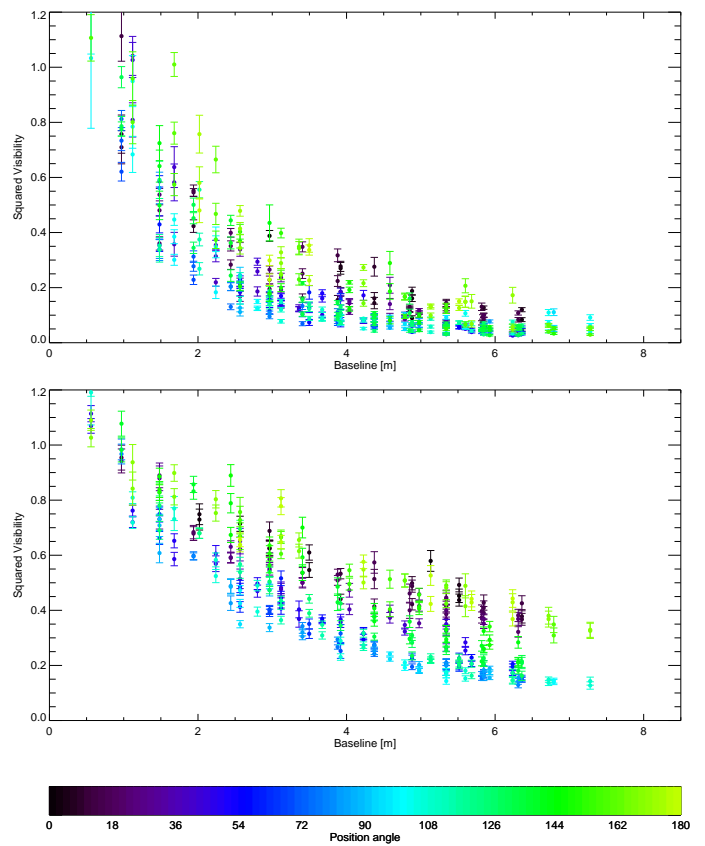}
  \caption{Calibrated $V^2$ of L$_{\rm 2}$ Pup at 2.30~$\rm\mu m$ (top) and 3.74~$\rm\mu m$ (bottom). Each visibility point is color-coded according to the respective baseline position angle.}
  \label{fig:viz}
\end{figure}

In the SAM power spectra (Fig.~\ref{fig:viz}), a dependence of the visibility on the baseline position angle (P.A.) is shown at similar baseline lengths for both wavelengths. The distinction is clearer at 3.74~$\rm\mu m$, where the difference for visibilities at P.A.$\le90$\degr and P.A.$\ge160$\degr is $V^2\sim0.2$ . 

For an initial determination of the overall extent of the detected structures, a simple source morphology was assumed (centro-symmetric) and geometric models were directly fitted to the Fourier data. The initial sizes were estimated by applying a circular Gaussian distribution fit to the azimuthally averaged visibilities, and the errors were estimated from the scatter of the azimuthally averaged measurements. None of these fits were satisfactory since the dependence of the visibility on the baseline P.A. was not taken into account. A further examination of the interferometric data was applied within a 20\degr\ cone, centered on the north--south and east--west directions. Tests were performed for both non-averaged and azimuthally averaged squared visibilities. When we applied new Gaussian distribution fits in all cases, the radius of the object was larger along the east--west orientation. A similar result is seen in the SAMPol data at 2.30~$\rm\mu m$. This preferential extension can also be noted in the \textit{uv}-planes of the original data (Fig.~\ref{fig:uvplane}).

\begin{figure}[!htbp]
	\centering
	\includegraphics[width=0.46\textwidth, bb=60 100 532 531]{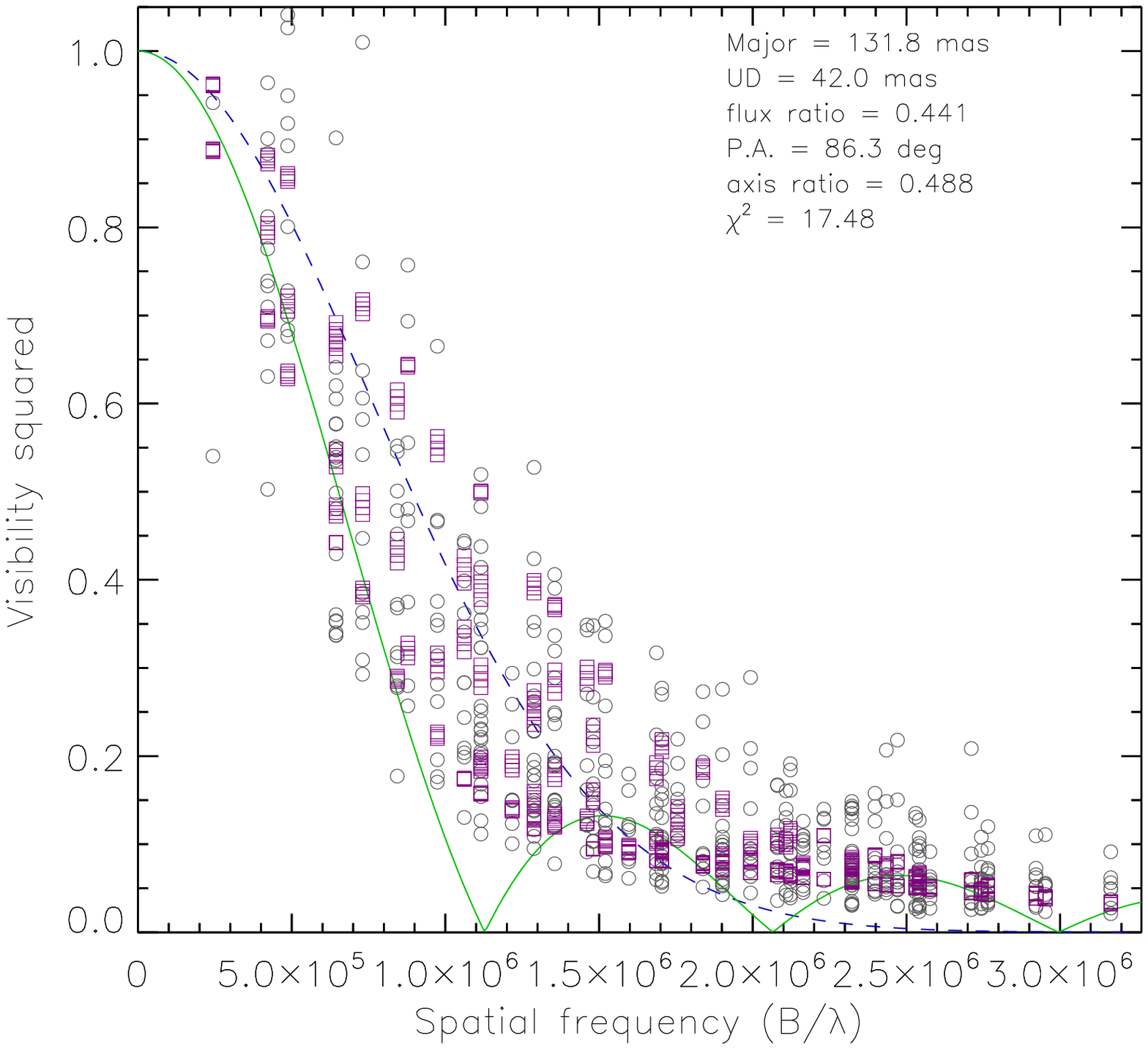}
	\includegraphics[width=0.48\textwidth, bb=60 152 552 583]{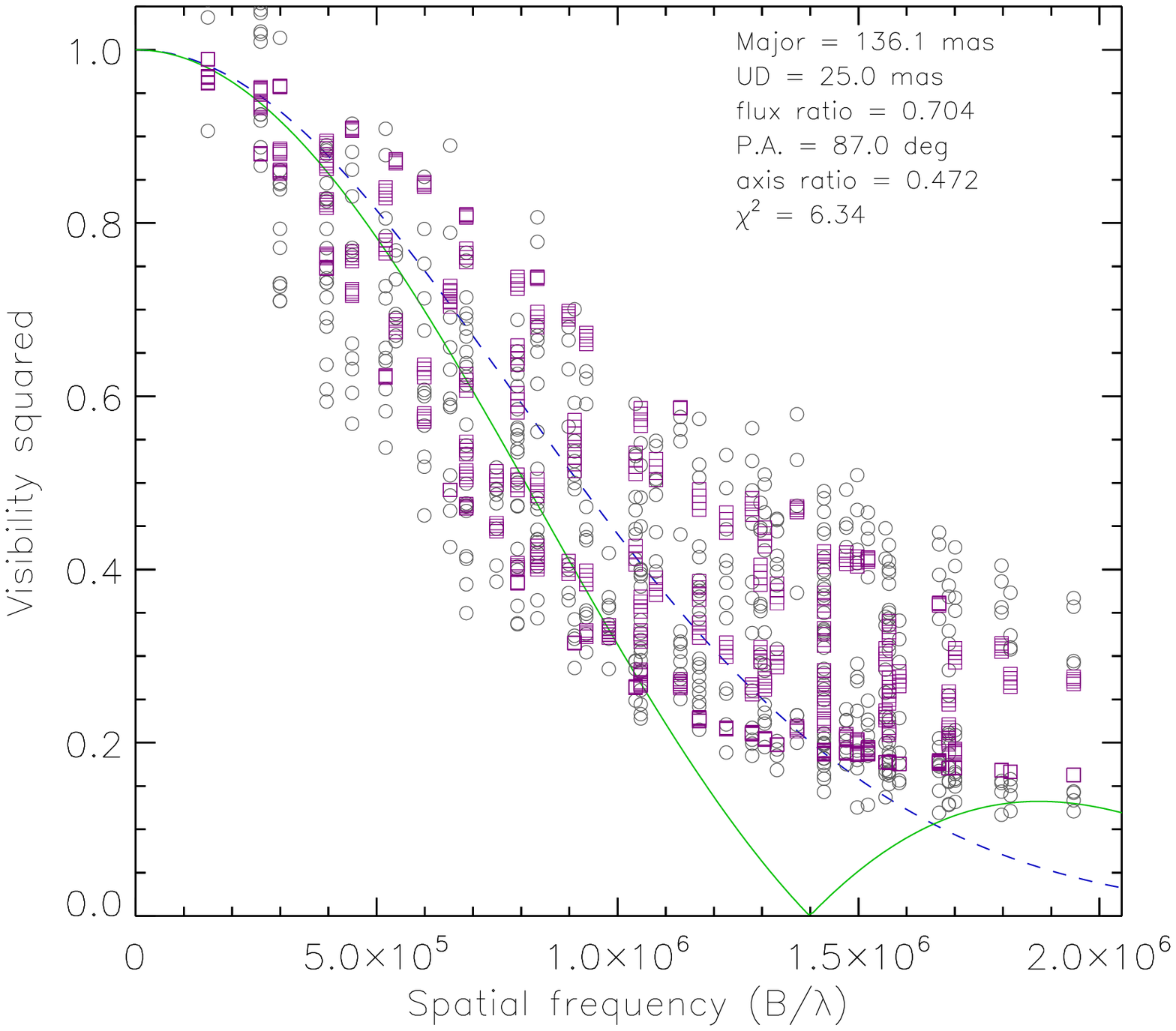}
	\caption{Best geometric model fit (uniform disk $+$ elliptical Gaussian; purple) to the SAM/NACO data (gray; as in Fig.~\ref{fig:viz}). The fitting results are given in the legend. The spread of the values at similar baseline lengths is due to the different baseline position angles. 2.30~$\rm\mu m$ and 3.74~$\rm\mu m$ data in the upper and lower panel.  Overplotted for comparison are the best fits of simpler, centrosymmetric functions, i.e., a uniform disk (green; lower line) and a circular Gaussian distribution (blue; middle dashed-line). The diameters ($\theta_{\rm UD}$) and FWHM of these fits are {\it upper panel} 223 and 102~mas, and {\it lower panel} 180 and 99~mas.}
	\label{fig:ellipse}
\end{figure}

Using the geometric model simulator \texttt{GEM-FIND} \citep{klotz2012}, we found that the shape of the object can be described by an ellipsoid, composed of a circular uniform disk representing the star and an elliptical Gaussian distribution representing the circumstellar environment. The best fits to the Fourier data are shown in Fig.~\ref{fig:ellipse}. Simpler centrosymmetric functions are also shown for comparison (a uniform disk and a Gaussian distribution).

The two-dimensional fitted structures are also overplotted for comparison to the NACO images in Fig.~\ref{fig:allimages}. The ellipsoid extends at approximately 140~mas along the major axis and at a similar position angle ($\sim$86\degr) in both wavelengths, while the size of the central component (uniform disk) varies from $42\pm0.3$ mas at 2.30~$\rm\mu m$ to $25\pm0.5$ mas at 3.74~$\rm\mu m$.

\begin{figure}
	\centering
		\includegraphics[bb=44 40 520 485,width=0.465\textwidth]{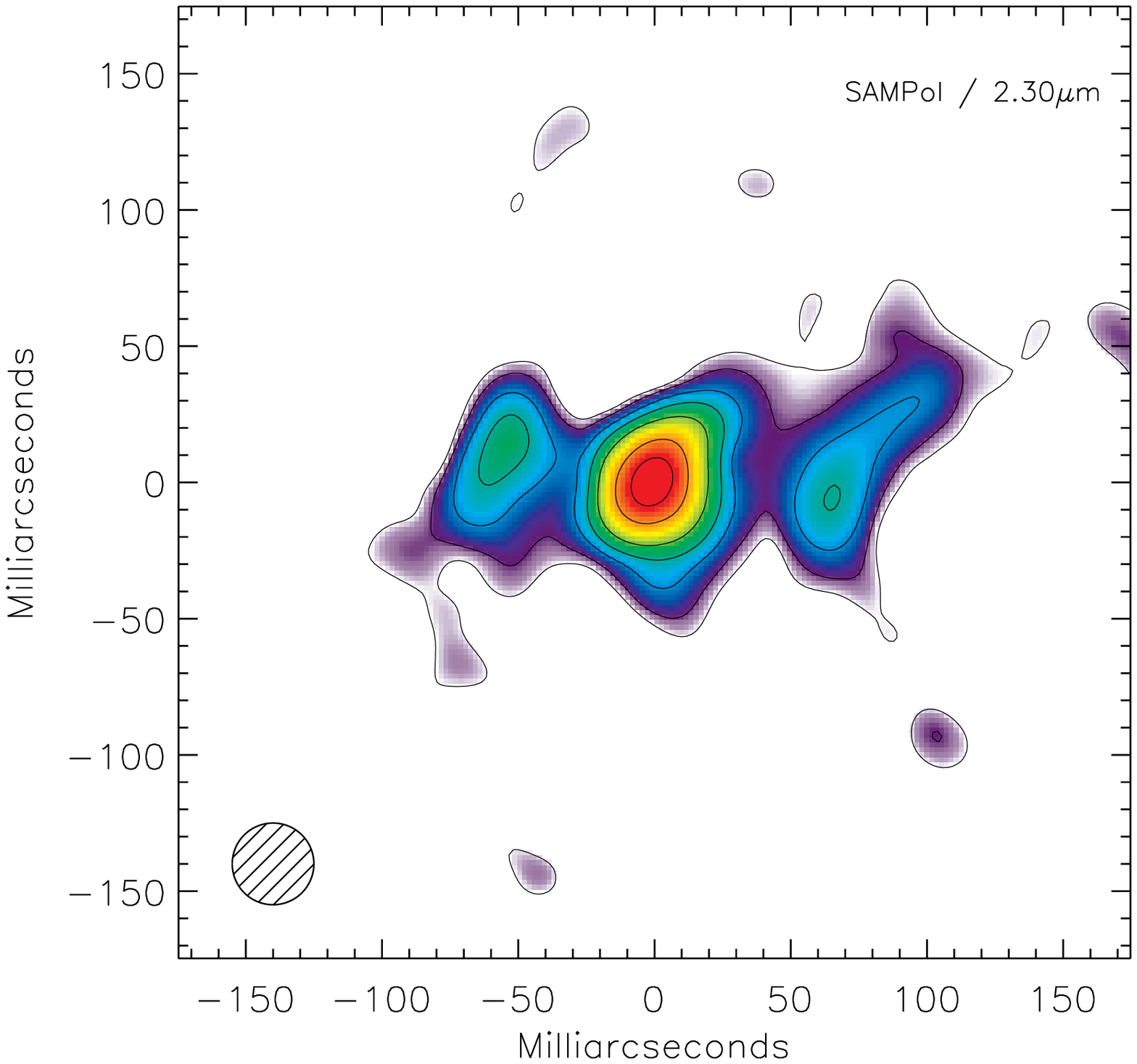}
	\caption{SAMPol Stokes $I$ map at 2.30~$\rm\mu m$ (February 2013) reconstructed with the BSMEM algorithm. Two clumpy structures emerge at $\sim$65 mas in the east--west direction at levels above 2\% of the peak intensity with a possible extension to the northeast. The central star (L$_{\rm 2}$ Pup) is unresolved at this wavelength ($\theta\leq30$ mas, $\geq50$\% peak intensity), but the surrounding envelope ($\theta\sim50$ mas) is non-spherical. Contour levels are at 1, 2, 5, 10, 20, 50, and 80\% of the peak intensity. North is up and east is on the right. }
	\label{fig:mapsFeb}
\end{figure}
\section{SAM/NACO imaging}\label{sec:imaging} 
Diffraction-limited maps were reconstructed from the Fourier observables using different methods to independently validate the results. These included the BiSpectrum Maximum Entropy Method~\citep[BSMEM; ][]{buscher1994}, the MArkov Chain IMager~\citep[MACIM; ][]{ireland2006}, and the Multi-aperture Image Reconstruction Algorithm \citep[MiRA; ][]{thiebaut2008}. Each of these methods uses a different algorithm and regularization techniques to produce the best-fit map from the interferometric observables (i.e., visibilities and closure phases).

\subsection{Image reconstruction with BSMEM}
The reconstruction algorithm BSMEM is based on the maximum entropy method~\citep[MEM;][]{sivia1987, skilling1984}. Based on the geometric model-fitting results (Section~\ref{sec:asymmetry}), images were reconstructed from the power spectra and the closure phases of the calibrated interferograms using a Gaussian distribution as an initial model. Although the algorithm converged at $\rm\chi^{2}\sim 1$, a small misfit was found at low spatial frequencies (that is, baselines shorter than one meter) between the final modeled power spectrum and the data at 2.30~$\rm\mu m$, suggesting either a miscalibration of the power spectrum at these spatial frequencies or that a more complex model was required.

Nevertheless, the image reconstruction produced similar structures, both in the SAMPol and SAM data in that wavelength (Figs.~\ref{fig:mapsFeb} and \ref{fig:allimages}). In the SAMPol Stokes $I$ map of February 2013, two bright {\it clumps} appear extended to $\sim$65 mas in the east--west direction at levels above 2\% of the peak intensity. In the SAM maps in particular, a bridge-type structure is more prominent in the west, while a hook-like pattern emerges in the east. It is unclear whether the circumstellar environment of L$_{\rm 2}$ Pup is polarized at 2.30~$\rm\mu m$. The Stokes $I$ map shown here was produced by conventional calibration, and this method could not provide more information on the source's polarization. However, when using a differential calibration method for the SAMPol data \citep{norris2012}, the method is sensitive to subtle polarization features. A more detailed analysis of the polarimetric data will be presented in a future paper.

A similar approach was followed for the 3.74~$\rm\mu m$ data. Here, no other misfits to the power spectra were found within the errors. In the maps presented in Fig.~\ref{fig:allimages}, we detect an elliptical structure with sizes 136 mas and 70 mas along the major and minor axes. This structure is also aligned in the east--west direction. At levels below 5\% of the peak intensity, two hook-like patterns are seen in the southwest and northeast directions.

Extended emission might be resolved out by the instrument, and/or the stellar photosphere is unresolved at a given wavelength. We emphasize that the MEM process selects the smoothest map that fits the original data, based on positivity and finite-extension of the map \citep{narayan1986,skilling1984}, and it is thus possible that another solution may have fitted the data better. MEM is also known to smooth out fine structures, for instance embedded point sources \citep{tuthill2002}. For these reasons, imaging results were reproduced with other different algorithms, as shown in the following sections.

\subsection{Image reconstruction with MACIM}
To verify the consistency of the previous reconstructions, we used MACIM \citep{ireland2006}. This algorithm uses a regularization method based on simulated annealing~\citep[for a detailed description, see][]{ireland2006}. The initial model consisted of a uniform disk of 25 mas in diameter\footnote{Based on the MAPPIT, SAM/NACO and VINCI diameters.} at 10\% of the total flux. The results are strikingly similar to those of BSMEM (\ga 1\% of the peak intensity), as shown in Fig.~\ref{fig:allimages}.

\subsection{Image reconstruction with MiRA}
\begin{figure*}[!ht]
  \centering
    \includegraphics[angle=90,bb=52 53 552 787,width=0.9\textwidth]{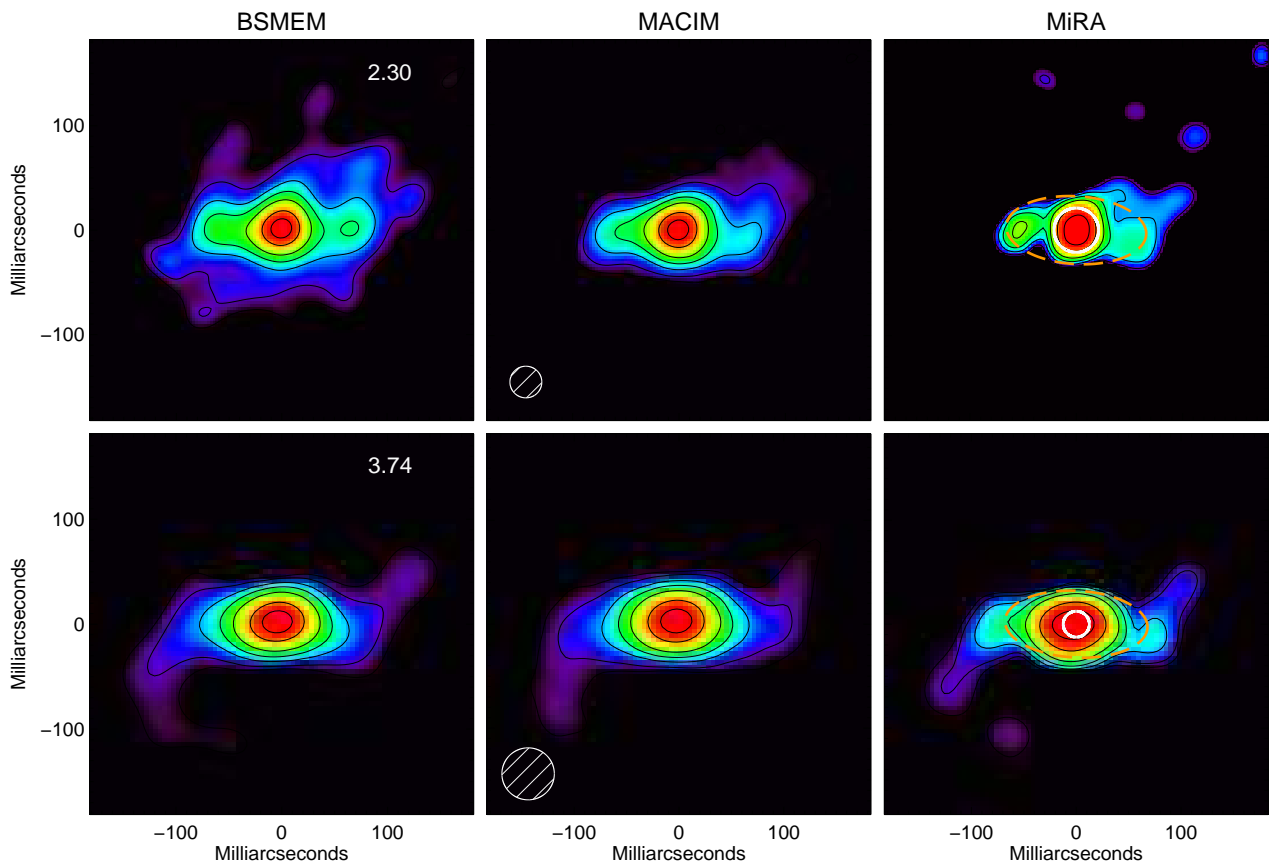}
  \caption{Comparison of image reconstruction for SAM maps at 2.30~$\rm\mu m$ (top) and 3.74~$\rm\mu m$ (bottom) from December 2012. Each column, from left to right, represents image reconstructions with BSMEM, MACIM, and MiRA. Contour levels and orientation are as in Fig.~\ref{fig:mapsFeb}. The resolution elements of SAM for the respective wavelengths are shown in the lower left corner of the second column. \texttt{GEM-FIND} model fits are overplotted in the third column for comparison: elliptical Gaussian (FWHM; orange) and circular uniform disk (white). The extent and size of the fitted models is similar to those of the mapped structures down to 5\% of the peak intensity, while $\geq$70\% of the emission originates from the central star. At 2.30~$\rm\mu m$ two clumpy patterns emerge in the east--west direction with a possible hook-like pattern rising in the north--east. The former are unresolved at 3.74~$\rm\mu m$, and the shape now resembles a bar with two hook-like patterns emerging in the north--east and south--west directions.}
  \label{fig:allimages}
\end{figure*}

MiRA is an algorithm that minimizes a penalty function composed of two terms following a gradient descent method \citep{thiebaut2008}. The two terms are the distance to the data ($\chi^2$ statistics) and the distance to some priors (such as `regularization', which is used to fill the Fourier space between the points of the {\it uv}-plane). MiRA implements many different regularizations; here we used the total variation with edge-preserving smoothness with an adopted regularization weight $\mu = 10^{3}$ \citep[for a detailed description, see ][]{thiebaut2008}. This regularization minimizes the total gradients in the image and subsequently smooths the flux levels in the image. The BSMEM and MACIM images are similar to MiRA, which succeeds in reproducing the overall shape of the structures, including the clumpy and hook-like patterns seen in all reconstructions (Fig.~\ref{fig:allimages}).

\subsection{Comparing image reconstruction and GEM-FIND fitting results}
In the right column of Fig.~\ref{fig:allimages}, the image reconstruction results are compared to the best-fit models of {\tt GEM-FIND}. While improvements can be implemented, the models provide reasonable\footnote{Although the reduced $\chi^2$ statistic of each model fit exceeds unity, these were the best fits found.} fits to the reconstructed images (reduced $\chi^2$ is 17.48 and 6.34 for 2.30~$\rm\mu m$ and 3.74~$\rm\mu m$, respectively; Fig.~\ref{fig:ellipse}):
\paragraph{Circular uniform disk:} the flux ratio of the central star over the circumstellar envelope is found to be 0.441 and 0.704 at 2.30 and 3.74~$\rm\mu m$. The respective modeled stellar diameters coincide with contour levels at 50 and 80\% of the peak intensity in the maps.
\paragraph{Elliptical Gaussian:} the major and minor axes, including the position angle of the ellipse, fit the structures in the maps to within 5\% of the peak intensity (cf. contours in Fig.~\ref{fig:allimages}) for both wavelengths.

\subsection{Comparison to Kervella et al. 2014}\label{ker}

The observations presented here are taken with the same instrument as was used in \citet{kervella2014}, but with a very different technique. Kervella et al. used direct imaging with adaptive optics as well as lucky imaging with very short integrations, selecting the best images. Deconvolution was used to further improve the resolution. This gives superb resolution but depends on the accuracy of the point-spread-function (PSF) calibration and on the Strehl ratio. The current data were taken with an aperture mask. This creates a Fizeau inteferometer, giving an exactly known PSF that is independent of the seeing (the seeing does affect the signal-to-noise), but with reduced throughput because of the mask geometry. 

Comparing the two techniques, aperture masking has higher angular resolution: its PSF is somewhat smaller than the diffraction limit of a single mirror and deconvolves well because it is mathematically known. Both techniques require accurate calibration (PSF for direct imaging and visibility amplitude for aperture masking). Aperture masking filters out extended emission, while adaptive optics leaves extended pedestals in the PSF. The techniques are complementary in view of their very different strengths.

Comparing our Fig. \ref{fig:mapsFeb} with Fig.~4 of \citet{kervella2014} illustrates this. The current data have a slightly better resolution and show smaller details, but the extended disk is better seen in Kervella et al. Allowing for this, the agreement is very good. The aperture masking $K$-band image shows two clumps either side of the star, indicating a better resolution than Kervella et al., but the extended disk emission is suppressed. The $L$-band image, with its shorter $uv$-spacings, shows the disk better. The overall structures agree very well, validating the images of \citet{kervella2014} and supporting the reliability of the image reconstruction.

\section{Nature of the detected structure}\label{sec:discussion} 

The NACO observations are focused at 2.30~$\rm\mu m$, where the detection of carbon monoxide emission\footnote{To our knowledge, there are currently no spectroscopic observations of the CO first-overtone band available in the literature.} and scattered dust emission from the circumstellar envelope is expected, and at 3.74~$\rm\mu m$, where continuum emission from the dusty environment probably is more prominent. The detected ellipsoidal structure is common in both wavelengths. This was confirmed by convolving the 2.30~$\rm\mu m$ maps with the resolution beam of SAM/NACO at 3.74~$\rm\mu m$. The overall structure may be interpreted either as an elongated outflow or a toroidal structure, depending on the inclination of the object with respect to the line of sight and the stellar rotation axis. The interpretation of the hook-like features depends on the inclination of the object as well. A few possible scenarios on the origin of the structure detected around L$_{\rm 2}$ Pup by SAM/NACO are presented below.

\subsection{Circumstellar shell hypothesis}

The structure detected at 2.30~$\rm\mu m$ is aspherical and extends on both sides to 65 mas on average (above 3\% of the peak flux east--west), which in physical units translates in 4.2 au. The structure becomes an ellipsoid at the pseudo-continuum (3.74~$\rm\mu m$), with approximate sizes 140~mas wide in the east--west direction and 70~mas wide in the north--south direction (Fig.~\ref{fig:allimages}). 

\citet{bedding2002} compared the IRTS mid-infrared spectrum (recorded in 1995, pre-dimming) to a model of a silicate-rich, detached shell\footnote{Nevertheless, they advised to use the best-fitting parameters of their model with caution.}. The inner and outer radii of that dust shell were $7.2\times10^{14}$ and $1.3\times10^{15}$ cm (or 48.1 and 86.9 au). At a distance of 64 pc, these dimensions translate into 0\farcs75 and 1\farcs35, and are placed within the extended envelope detected at 11.7~$\rm\mu m$ \citep[1\farcs8$\times$1\farcs8;][]{jura2002}. With a known expansion velocity\footnote{This is the outflow velocity of gas, not dust. These are probably only loosely coupled in a star with such a low mass-loss rate \citep[$\sim3\times10^{-8}\rm M_{\sun}yr^{-1}$; ][]{schoier2004}.} of approximately 2.5 km s$^{-1}$\citep{kerschbaum1999}, the outer edge of the shell would have reached that distance within 165 years. Nevertheless, such a shell is much larger than the structures detected by SAM and also lies outside the region imaged by \citet{kervella2014}.

There have been previous detections of molecular shells in the extended atmospheres of AGB stars~\citep{lebouquin2009, perrin2004, eisner2007, thompson2002, ohnaka2012, wittkowski2007, wittkowski2011, ireland2004c}. The sizes of these shells have been estimated through star+molecular shell models (MOLspheres) or dynamic atmosphere models and are spherically symmetric. The observations presented here showed that the case of L$_{\rm 2}$ Pup is different because no spherically-symmetric shell was detected. 

\subsection{Toroidal structure hypothesis}\label{sec:tor}

The molecular emission of L$_{\rm 2}$ Pup has been extensively monitored in the past fifty years; it showed the variability of the outflow velocity~\citep{kerschbaum1996,kerschbaum1999,winters2002,lepine1976, winters2002, winters2003, haikala1990, haikala1994, menten1991, gomezbalboa1986, lepine1978, caswell1971, knowles1978, balister1977, schoier2004, gonzalez-delgado2003}. \citet{gonzalez-delgado2003} reported that the interior of the circumstellar envelope of L$_{\rm 2}$ Pup is probably complex, based on their SiO observations. However, the radio interferometric measurements by \citet{schoier2004} of ``thermal SiO'' were unable to probe the direction of these flows, and a circularly symmetric envelope was assumed.  

Recent observations \citep{mcintosh2013b} of the variation in the positioning of the velocity centroids of SiO maser emission over a period of four years, indicate an asymmetric nature of the gaseous environment of L$_{\rm 2}$ Pup which may be associated with a rotating edge-on disk and bipolar ejecta (see Sect.~\ref{maser}). SiO maser emission probes the material much closer to the star than the ``thermal'' SiO emission. \citet{kerschbaum1999} suggested that the line profile seen in the carbon monoxide spectra is caused by a gaseous disk around the star, similar to the case of SV Psc \citep{klotz2012b}. \citet{bujarrabal2013} showed that similar CO line profiles can be found in post-AGB stars, and they assigned these to rotation in molecular disks. In fact, three of these objects -- the Red Rectangle, AC Her, and HR 4049 -- have slow expansion velocities, bipolar ejecta, and circumbinary dusty disks.

The rotational axis of L$_{\rm 2}$~Pup is unknown, which allows for a range of scenarii on the nature of the detected structures. However, the images of \citet{kervella2014} are strongly suggestive of a rotating east-west disk, which would indicate a north-south rotation axis.

Considering that the bilateral clumps are observed farther out than the extended circumstellar emission (\ga30~mas) in all 2.30~$\rm\mu m$ maps, but are absent from the 3.74~$\rm\mu m$ maps\footnote{Due to the lower resolution at the wavelength.}, it is possible that the edge-on structure detected with SAM/NACO is in fact a torus of gas interwoven with dust. The origin of the hook-like patterns is quite perplexing. These emerge at levels $\leq$5\% of the peak intensity and extend symmetrically southwest and northeast in all the SAM maps. Assuming that the inclination of L$_{\rm 2}$~Pup in our line of sight is perpendicular to the north--south axis, that is $90\degr$, the possibility of an inclined, edge-on spiral or helical structure, similar to those observed in AGB stars \citep[e.g., AFGL3068;][]{mauron2006} and planetary nebulae \citep[e.g., PNG356.8+03.3;][]{sahai2011}, should not be excluded.

\subsection{Stellar and torus parameters}\label{maser}

The SiO maser spectra of \citet{mcintosh2013b} show an unusual profile, with two peaks separated by 20 km s$^{-1}$, in addition to the more usual narrow peak at the systemic velocity. A rotating gaseous disk would fit this profile, where the separated peaks show the approaching and receding side of the disk. The central velocity component may trace the stellar wind might be emission from the disk in front of the star, where the maser is amplifying the stellar emission. This model gives a rotational velocity of the disk of $v \sin i = 10\,\rm km\,s^{-1}$, where $i$ is the inclination with respect to the plane of the sky. 

The two clumps seen on either side of the star are consistent with the expected structure seen at the inner radius of the disk. This gives an inner radius of 4.2\,au (Kervella et al. derived 6\,au from their models). 

Assuming Keplerian velocities, this gives a mass of the star of $M \sin i = 0.7\,\rm M_\odot$. The optical obscuration events, and the images of \citet{kervella2014}, indicate an inclination $i \sim 90\degr$, although given the size of the star, a significant range of $i$ is possible. However, using a maximum range of $i=60\deg$ and $r=6\,au$ still implies a stellar mass $M <1\,\rm M_\odot$. 

This mass is lower than that derived by Kervella et al. (2\,M$_\odot$). They located the star at the beginning of the AGB (meant is the start of the TP-AGB) where the mass is still close to the initial mass. However, the short period of the star of 140 days  together with a location 0.5\,mag below the Mira PL relation \citep{bedding1998, bedding2002} suggest a low initial mass, as found here.

\subsection{Binarity hypothesis}

The possible signature of a binary companion was investigated in the SAM/NACO data, although the binary nature of L$_{\rm 2}$ Pup was previously doubted by \citet{jorissen2009}. Using the geometric model-fitting tool LITpro\footnote{LITpro software available at http://www.jmmc.fr/litpro}~\citep{litpro}, an additional, unresolved component, with a flux ratio ranging from 1/100 to 1/500, was included in the model presented in Sect.~\ref{sec:asymmetry} and was positioned randomly within the NACO field of view. All attempts failed to converge ($\chi^2>>1$), suggesting that no signal from a binary companion was found in the data. 

\section{Conclusions}\label{sec:results} 
The SAM and SAMPol diffraction-limited imaging revealed the complex circumstellar nature of the envelope of L$_{\rm 2}$ Pup in both gaseous emission and dust continuum. The maps at 2.30~$\rm\mu m$ show an asymmetric structure extending to $\sim$65 mas in the west at levels above 5\% of the peak intensity and a hook-like extension (Figs.~\ref{fig:mapsFeb} and \ref{fig:allimages}). It resembles a linear structure  from the star, whose extended envelope is resolved at $\sim$30 mas, although the stellar photosphere should be smaller than that. A fainter, hook-like feature emerges in the northeastern direction. The overall shape of the structure is similar at 3.74~$\rm\mu m$, where it resembles an ellipsoid with sizes 140 mas and 70 mas along the major and minor axis, respectively. This warm dust component extends farther out into a fainter, helical feature.

Both structures are oriented in the east--west axis, agreeing with \citet{kervella2014}, but different from previous aperture-masking observations: \citet{ireland2004} positioned their mask in a north--south orientation (P.A.=180\degr) and failed to detect the equatorial structure.

This ellipsoidal structure, found in pseudo-continuum (3.74~$\rm\mu m$), is the less resolved counterpart of the more detailed, bilateral structure (2.30~$\rm\mu m$), since the dimensions of both are similar. The structure is contained within the extended silicate-rich dust zone \citep{bedding2002}. As shown in the near-infrared maps, it appears that the structure detected in L$_{\rm 2}$ Pup is seen nearly \textit{edge on}.  We interpret the structure as a torus with inner radius 4.2\,au around L$_{\rm 2}$ Pup. This torus is probably the cause of the long-lived dimming event \citep{bedding2002}. It might be that a more extended dust reservoir is located between the torus and the extended envelope~\citep{jura2002}, although it is currently outside the field of view of SAM/NACO. 

SiO maser profiles indicate that the torus may be rotating at 10 km s$^{-1}$. Using the inner radius of 4.2 au, we derive a stellar mass of 0.7\,M$_\odot$ with an upper limit of 1\,M$_\odot$. This low mass is consistent with the short period of L$_{\rm 2}$ Pup.


\begin{acknowledgements}
We would like to thank the NACO operational team for their assistance during our observations, as well as K.~Hebden, W.~Nowotny and T.~Lebzelter for their valuable comments. FL, DK and JH were supported by the Austrian science fund FWF project number AP23006. CP was supported by the Belgian Federal Science Policy Office via the PRODEX Programme of ESA. JK was supported by the French ANR POLCA project (Processing of pOlychromatic interferometriC data for Astrophysics, ANR-10-BLAN-0511). We acknowledge with thanks the variable star observations from the \textit{AAVSO International Database} contributed by observers worldwide and used in this research. This research has made use of the Jean-Marie Mariotti Center \texttt{LITpro} service co-developed by CRAL, LAOG and FIZEAU.
\end{acknowledgements}

\bibliographystyle{aa}
\bibliography{refs}

\appendix
\section{Visual light curve}
\begin{landscape}
\begin{figure}[!htb]
 \centering
\includegraphics[angle=90, width=1.3\textwidth, bb=32 51 563 785]{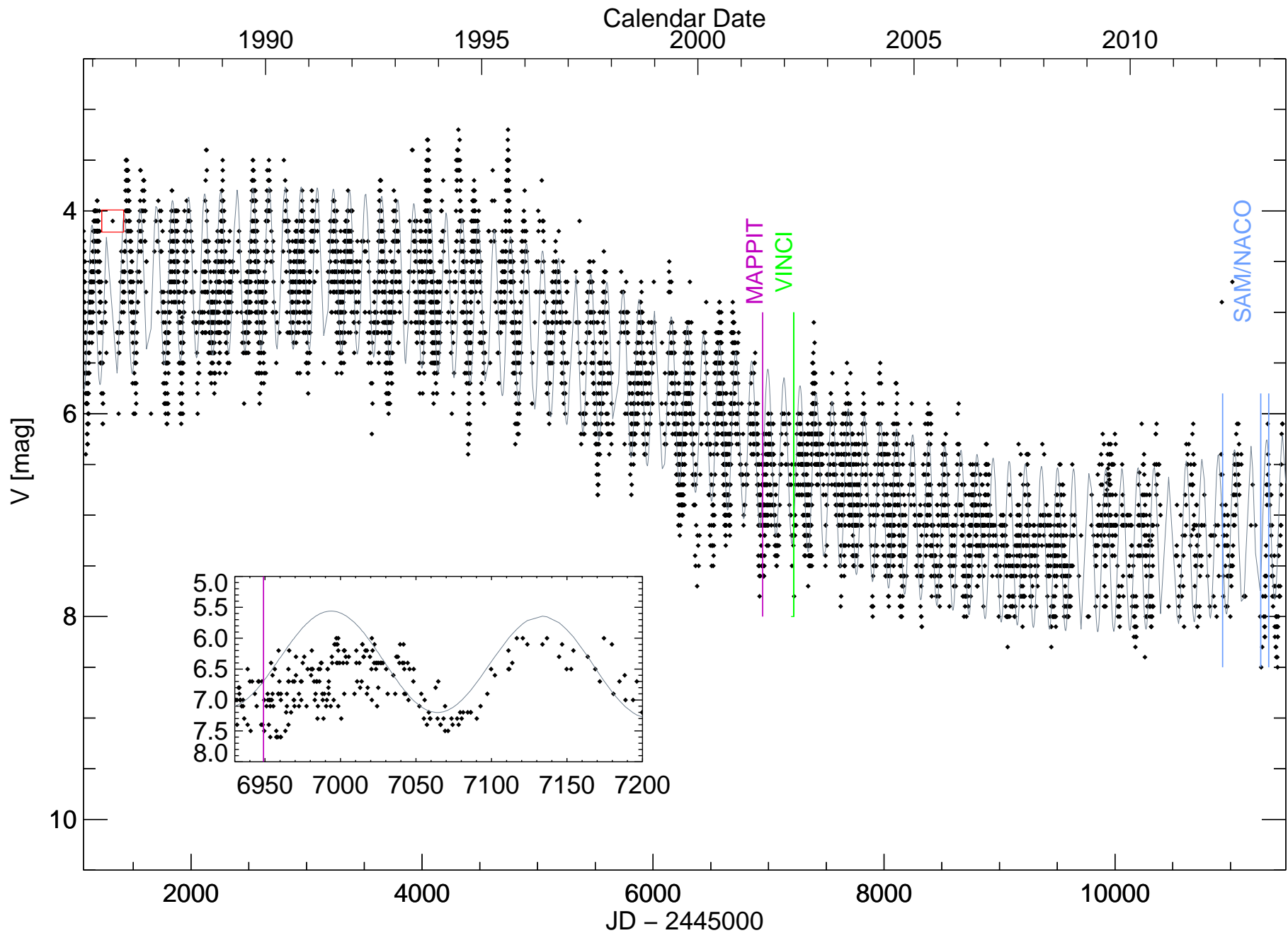}
\caption{Visual light curve of L$_{\rm 2}$ Pup from 3 January 1985 to 6 July 2013 (data from AAVSO). The dimming event started in 1994, and the stellar luminosity in the V band has not increased to pre-1994 values ever since. The observing periods for MAPPIT, VINCI, and NACO are marked in magenta, green, and blue lines, while the selected Julian date for zero pulsation phase, selected on JD=2\,446\,322, is drawn as a red box. An average light curve (gray) is overplotted to guide the eye. The inset shows a magnified version of the fitted light curve over two pulsation cycles, during and right after the MAPPIT observations. }
\label{fig:lightcurve}
\end{figure}
\end{landscape}

Figure~\ref{fig:lightcurve} shows the visual light curve of L$_{\rm 2}$~Pup from 1985 to 2013 (data from AAVSO), where the observing periods of MAPPIT, VINCI, and NACO are clearly marked. An average light curve is overplotted to guide the eye.

\end{document}